\documentclass[aps,onecolumn]{revtex4}
\usepackage{graphicx}
\usepackage{epsfig}
\usepackage{epstopdf}
\usepackage{amsfonts}
\usepackage{amssymb}
\usepackage{amsbsy}
\usepackage{amsmath}
\usepackage{mathrsfs}
\usepackage{latexsym}
\usepackage{natbib}
\usepackage{bm}
\usepackage{color}
\usepackage[a4paper, total={6.5in, 9.8in}]{geometry}

\usepackage{braket}
\usepackage{slashed}
\usepackage{pgfplots}
\usepackage{natbib}
\usepackage{yfonts}
\usepackage{amsfonts}
\usepackage[T1]{fontenc}

\usepackage{tikz}
\usetikzlibrary{shapes,arrows,shadows}

\begin{document}

\title{Shadow of Non-singular Rotating Magnetic Monopole in Perfect Fluid Dark matter}

\author{Gowtham Sidharth M}
\email{gowthamsidharth.m2019@vitstudent.ac.in}

\affiliation{ School of Advanced Sciences\\
	Vellore Insitute of Technology\\
	Chennai, India 600048 }

 \author{Sanjit Das}
\email{sanjit.das@vit.ac.in}

\affiliation{ School of Advanced Sciences\\
	Vellore Insitute of Technology\\
	Chennai, India 600048 }

\date{\today}

\begin{abstract}
\begin{center}
\underline{\textbf{Abstract}}
\end{center}
Bardeen proposed a gravitationally collapsing magnetic monopole black hole solution that is free of singularity. In this article, we have studied the size and shape of the rotating Bardeen black hole shadow in presence of perfect fluid dark matter. we have discussed how parameters such a spin, magnetic monopole charge, and influence of dark matter affect the shadow of our black hole. the apparent shape of the black hole was studied by using two observables, the radius $R_s$ and the distortion parameter $\delta_s$. Further the blackhole emission rate is also studied, we found out that for rotating Bardeen in PFDM, For a constant monopole charge, the emission rate increases with increase in dark matter parameter and decreases with increase in spin parameter, and for constant dark matter parameter, the emission rate decreases with increase in magnetic charge and spin.
\end{abstract}

\maketitle

\section{Introduction}

From observations of gravitational waves to blackhole shadows recent astrophysical observations provided more and more conclusive evidence for the existence of blackholes\cite{Abbott2016} \cite{Akiyama2019}. Blackhole shadows are the result of strong gravitational lensing which for a distant observer views as a bright circle with a dark interior. Synge \cite{Synge:1966okc} showed that for a non-spinning spherically symmetric Schwarzschild type blackhole the shadow boundary would be a perfect circle. later Bardeen \cite{Bardeen}predicted that the rotating Kerr-like black holes will have a deformed circle. Recent studies have motivated authors to explore the theoretical aspects of Kerr-Newman blackhole,Kerr-Nut spacetimes \cite{Hioki:2009na} \cite{vries_1999},Further research in blackhole shadows in the background of higher dimensional and alternative theories of gravity where also carried \cite{Papnoi:2014aaa,Abdujabbarov:2015rqa,Singh:2017vfr,Amarilla:2010zq,Dastan:2016vhb,Kumar:2017tdw,Vetsov:2018eot}.

Blackholes are the purest of objects in nature. Its making requires only the concepts of space and time. Blackhole in its heart holds the most complex, yet beautiful set of equations that theoreticians have come across, a singularity. Noble laureate Dr.Roger Penrose claims in his famous cosmic censorship conjecture that all singularities must be dressed by an event horizon\cite{Senovilla:1998oua} \cite{Wald:1997wa}. The peculiarity of this conjecture is that even though it rules out the existence of naked singularity, it does not restrict the possibility Non - Singular blackholes from existing. This was discussed briefly by Hawking and Ellis \cite{hawking_ellis_1973}. In 1968 Bardeen proposed his non - singular blackhole model at the GR5 conference. Bardeen's model avoids strong energy condition and satisfies weak energy condition. Following the Bardeen model, many singularity-free blackholes are proposed \cite{Cabo:1997rm,Mars:1996khm,Barrabes:1995nk} but these regular Bardeen models do not arise as a direct solution for Einstein equations which means that they don't have any physical sources. Recent studies have associated some physical sources with these regular blackholes \cite{Ayon-Beato:1998hmi,Ayon-Beato:1999qin,Ayon-Beato:1999kuh}.

A physical interpretation for the Bardeen Blackhole was given by Eloy Ayon-Beato and Alberto Gracia where they associated the Bardeen blackhole to a self-gravitating magnetic monopole\cite{Ayon-Beato:2000mjt}. The following metric expresses the Bardeen model as,
\begin{equation}
ds^2 =-(1-\frac{2Mr^2}{(r^2+g^2)^{\frac{3}{2}}})dt^2 +(1-\frac{2Mr^2}{(r^2+g^2)^{\frac{3}{2}}})^{-1} dr^2+r^2 d\Omega^2
\end{equation}
where the $g$ is the monopole charge that arises from non-linear electrodynamics.

Recent observational data shows that our universe is expanding and this rate of expansion is indeed accelerating \cite{Albrecht:2006um}\cite{SupernovaSearchTeam:1998fmf}. This could mean that there is a presence of negative pressure. There are two ways which could explain the presence of negative pressure, one the cosmological constant and secondly, the quintessential dark energy \cite{Frieman:2008sn}. A review on cosmological constant is given in \cite{Carroll:2000fy}In recent years many authors studied the theoretical aspects of black holes surrounded by quintessence\cite{Singh:2017xle,Abdujabbarov:2015pqp,Zeng:2020vsj,Khan:2020ngg,Pedraza:2020uuy,Toshmatov:2015npp}. In wake of this, different dark matter models have also gained attention.Blackhole shadows in cold dark matter(CDM)\cite{Dubinski:1991bm}\cite{1996ApJ} and scalar field dark matter (SFDM) \cite{Spergel:1999mh}have been studied.

For our work we have choose an alternative dark matter model, one that was proposed by keislev \cite{Kiselev:2002dx}\cite{Kiselev:2003ah}where he described the dark matter as perfect fluid.The perfect fluid dark matter(PFDM)\cite{Rahaman:2010xs} can explain the asymptotical rotation of spiral galaxies. Recently many have studied the blackhole shadows in PFDM \cite{Atamurotov:2021hck,Ma:2020dhv,Haroon:2018ryd}, blackhole surrounded by PFDM immersed in plasma was studied in \cite{Atamurotov:2021hoq}\cite{Das:2021otl}. Horizon structure of Bardeen blackhole was discussed in \cite{zhang_chen_ma_he_deng_2021} . Futher blackhole shadows surrounded by PFDM in addition to cosmological constant and quintessence were also carried out\cite{Ndongmo:2021how}.

In this paper, we are going to study the shadow of rotating Bardeen black hole in Perfect Fluid dark matter. The paper is structured in the way that next section we would derive the first integrals for the rotating metric. In section 3 the effective potential is discussed. In section 4, we have calculated the shadow and observables. In section 5 we have studied the blackholes emission rate. Finally, in section 6 we conclude the paper with a brief discussion and interpretation.

\section{Rotating Gravitationally Collapsed Magnetic Monopole in Perfect Fluid Dark matter}

The metric of a rotating gravitationally collapsed magnetic monopole in PFDM is achieved by the Newman Janis algorithm \cite{zhang_chen_ma_he_deng_2021},\\

\begin{equation}\label{e2}
ds^2 = -(1-\frac{2\rho r}{\Sigma})dt^2 + \frac{\Sigma}{\Delta_r}dr^2+\Sigma d\theta^2 -\frac{4 a \rho r \sin^2\theta}{\Sigma}dtd\phi -\sin^2\theta(r^2+a^2+\frac{2 a^2 \rho r \sin^2 \theta }{\sigma})d\phi^2
\end{equation}
with \\
\begin{equation}
2 \rho = \frac{2Mr^3}{(r^2+g^2)^{\frac{3}{2}}} -k \ln\frac{r}{|k|},
\end{equation}
\begin{equation}
\Sigma = r^2+a^2 cos^2\theta
\end{equation}
\begin{equation}
\Delta_r = r^2+ a^2 -\frac{2Mr^3}{(r^2+g^2)^{\frac{3}{2}}} + k r \ln\frac{r}{|k|}
\end{equation}
\\

The equations for photon orbit is derived using Hamilton-Jacobi variable seperation method. The Hamilton-Jacobi equation in its general form expressed as\\
\begin{equation}\label{h1}
\frac{\partial S}{\partial \lambda} = -\frac{1}{2}g^{\mu\nu}\frac{\partial S}{\partial x^{\mu}} \frac{\partial S}{\partial x^{\nu}}
\end{equation}
where $\lambda$ is the affine parameter and S corresponds to the Jacobi action\\
\begin{equation}\label{h2}
S = \frac{1}{2}m^2 \lambda - E t + L \phi + S_r(r)+S_{\theta}(\theta)
\end{equation}
where m is the rest mass, E and L are energy and Angular momentum which are the constants of motion.

with \ref{h1} and \ref{h2}, one arrives at
\begin{equation}
\Delta_r (\frac{\partial S_r}{\partial r})^2 + (\frac{\partial S_{\theta}}{\partial \theta})^2 + \frac{L^2}{Sin^2 \theta} - E^2 a^2 Sin^2\theta - \frac{1}{\Delta_r} (a^2L^2+E^2(r^2+a^2)^2-2a L E (r^2+a^2))+\frac{2 a E L(2 r \rho - r^2-a^2)}{\Delta_r} = 0
\end{equation}
Following\cite{1983mtbh.book.....C} the solution for $S_r$ and $S_\theta$ yeilds,
\begin{equation}
\Delta_r (\frac{\partial S_r}{\partial r})^2 = (a L -(a^2+r^2)E)^2-(\mathcal{K}+(a E-L)^2)\Delta_r = \mathcal{R}(r)
\end{equation}
\begin{equation}
\frac{\partial S_r}{\partial r} = \frac{\sqrt{\mathcal{R}(r)}}{\Delta_r}
\end{equation}
\begin{equation}
(\frac{\partial S_{\theta}}{\partial \theta})^2 = \mathcal{K} - (\frac{L^2}{Sin^2\theta}-a^2E^2)Cos\theta = \Theta(\theta)
\end{equation}
\begin{equation}
\frac{\partial S_{\theta}}{\partial \theta} =\sqrt{\Theta(\theta)}
\end{equation}
where $\mathcal{K}$ is the the seperation constant.
The trajectory of a photon is calculated with two impact parameter,
\begin{equation}
\xi =\frac{L}{E} ,\eta = \frac{\mathcal{K}}{E^2}
\end{equation}
Rewritting $\mathcal{R}(r)$ and $\Theta(\theta)$in terms of impact parameters,
\begin{equation}\label{r1}
\mathcal{R}_p(r) = (a \xi -(a^2+r^2))^2-(\eta +(a-xi)^2)
\end{equation}
\begin{equation}
\Theta_p = \eta -(\frac{\xi^2}{Sin^2\theta} - a^2)cos^2\theta
\end{equation}
where,
\begin{equation}
\mathcal{R}_p (r)=\frac{\mathcal{R}}{E^2} ,\Theta_p(\theta) =\frac{\Theta(\theta)}{E^2}
\end{equation}

The equation of geodesic motion can be written the form \cite{Hou2018}
\begin{equation}\label{s1}
\Sigma \dot{t} = \frac{((r^2+a^2)E-aL)(r^2+a^2)}{\Delta_r}-a(aE Sin^2(\theta)-L)
\end{equation}
\begin{equation}
\Sigma\dot{r} =\sqrt{\mathcal{R}(r)}
\end{equation}
\begin{equation}
\Sigma \dot{\theta} =\sqrt{\Theta(\theta)}
\end{equation}
\begin{equation}\label{s2}
\Sigma \dot{\phi} = \frac{a((r^2+a^2)E-aL)}{\Delta_r}-\frac{aE Sin^2(\theta)-L}{Sin^2(\theta)}
\end{equation}

\section{Effective Potential}
A photon with maximum potential energy will have null radial velocity and null radial acceleration, satisfying the condition for spherical orbit. the stability of the orbit is determined by the effective potential where the stable orbits have effective potential minima and unstable orbits have effective potential maxima. the effective potential is defined as,
\begin{equation}
V_{eff} = \frac{E^2 - 1}{2} - \frac{1}{2}\dot{r}^2
\end{equation}
where $\dot{r} =\frac{\sqrt{\mathcal{R}}}{\Sigma}$, the $\dot{r}$ is the derivative with respect to affine parameter $\lambda$

Looking at fig.1 we can see that the difference in maxima and minima of effective potential decreases with an increase in magnetic moment g value.
\begin{figure*}
\centering
\includegraphics{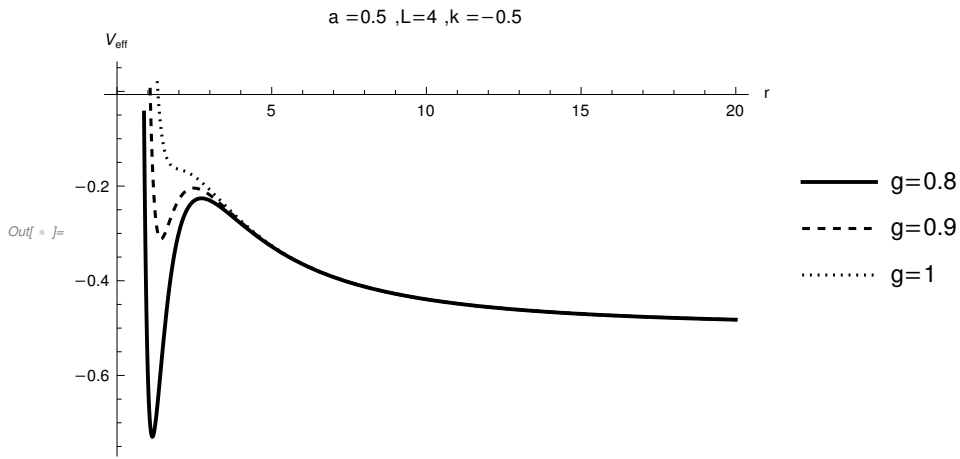}
\caption{Examples of Effective potential of rotating Non-singular Magnetic Monopole in Perfect Fluid Dark matter for different values of magnetic monopole value g with a = 0.5, L=4 and k = -0.5}
\end{figure*}

\begin{figure*}
\centering
\includegraphics{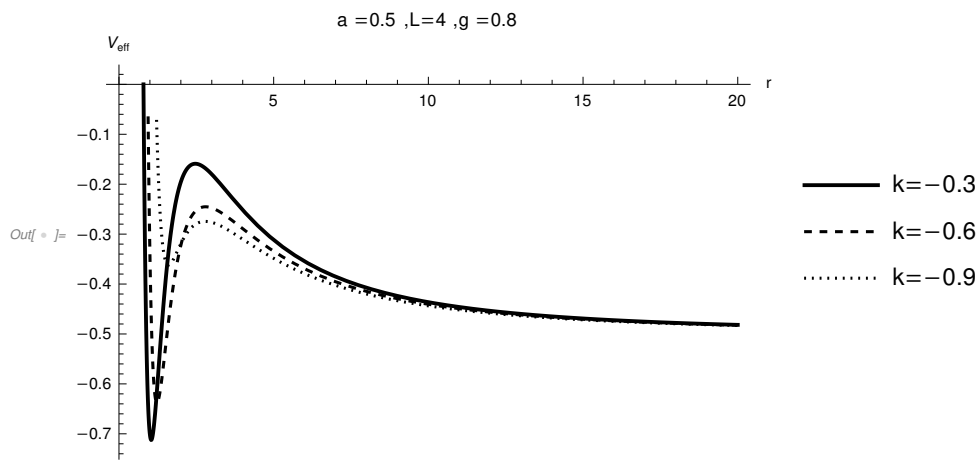}
\caption{Examples of Effective potential of rotating Non-singular Magnetic Monopole in Perfect Fluid Dark matter for different values of dak matter parameter k with a = 0.5,L=4 and g = 0.8}
\end{figure*}
for a photon to have spherical orbit,it should have null radial velocity and null radial acceleration,which means that,
\begin{equation}\label{r2}
\mathcal{R}_p(r) =0, \frac{d \mathcal{R}_p(r)}{d r} =0
\end{equation}
Using \ref{r1} and \ref{r2}, we can write $\xi$ and $\eta$ as
\begin{equation}
\xi = \frac{- 4 r \Delta_r + a^2\Delta'_r + r^2 \Delta'_r}{a \Delta'_r}
\end{equation}
\begin{equation}
\eta = \frac{r^2(16 a^2\Delta_r-16\Delta^2_r+8r\Delta_r\Delta'_r-r^2\Delta'^2_r)}{a^2 \Delta'^2_r}
\end{equation}
\section{Shadows}
In this section, we investigate the shadow of gravitationally collapsed rotating magnetic monopole in the presence of perfect fluid dark matter. The Shape of the black hole depends on the celestial coordinates $\alpha$ and $\beta$, which is given as

\begin{equation}
\alpha = \lim_{r_0 \rightarrow \infty}(-r^2_0 Sin\theta \frac{d \phi}{d r}|\theta \rightarrow i)
\end{equation}
\begin{equation}
\beta = \lim_{r_0 \rightarrow \infty}(-r^2_0 \frac{d \theta}{d r}|\theta \rightarrow i)
\end{equation}
where $r_0$ is the distance between the observer and blachole and i is the angle between blackhole's rotational axis and observer's line of sight.The celestial coordinates can be expressed as a funtion of $\xi$ and $\eta$ with the help of geodesic equations (\ref{s1}) - (\ref{s2}),
\begin{equation}
\alpha = \frac{\xi}{Sin( i)}
\end{equation}
\begin{equation}
\beta = \pm \sqrt{\eta+a^2 Cos^2 (i) -\xi^2 cot^2( i) }
\end{equation}
In equatorial plane,the above equations reduces as
\begin{equation}
\alpha = -\xi
\end{equation}
\begin{equation}
\beta =\pm \sqrt{\eta}
\end{equation}

Fig 2 shows the different shadows obtained by varying dark matter parameter \textfrak{d} for different values of spin parameter a and magnetic moment g, it is clear that from the fig that the size of the shadow increases with a decrease in magnetic moment g, the shadows are a perfect circle for the non-rotating case(i.e a=0). As a increases, the distortion produced due to spin is also more prominent for the lower values of k. Fig 3 shows the shadows for varying magnetic moment g with different values of a and k. Although the size of the shadow does not exhibits any major changes , the distortion caused by a is prominent for higher values of g.

\begin{figure}[h]
\centering
\includegraphics[scale=0.75]{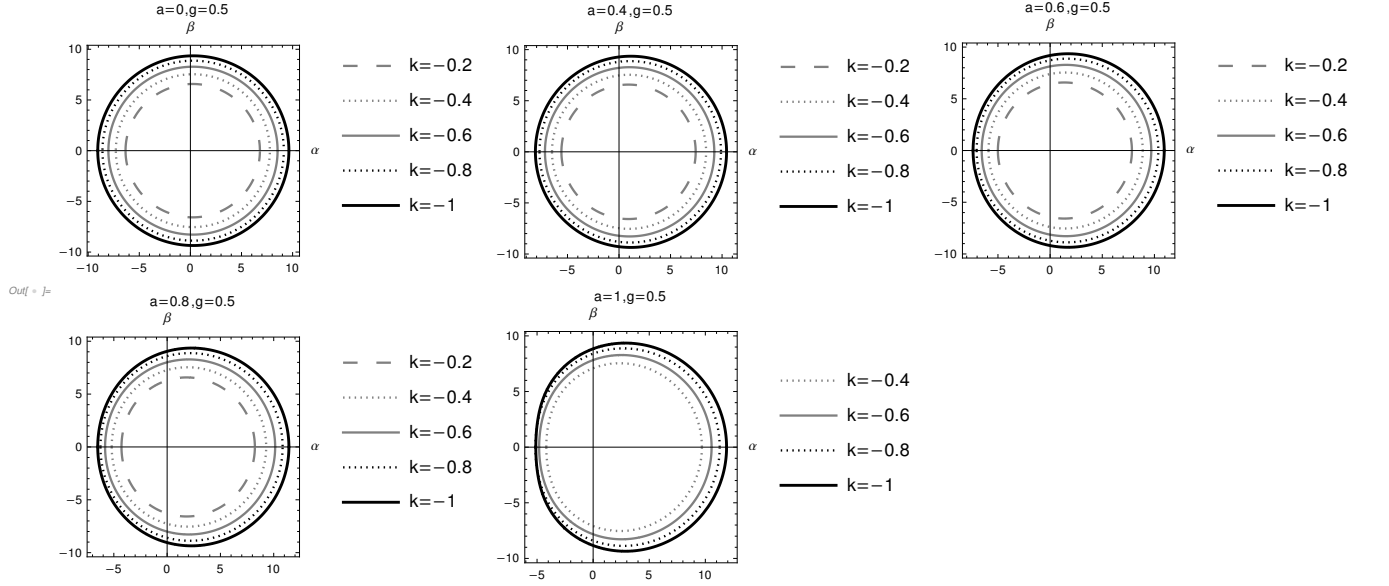}
\caption{Shadows of gravitationally collapsed magnetic monopole in PFDM for varying dark matter parameter k for different values of spin parameter a and magnetic moment g }
\end{figure}

\begin{figure}[h]
\centering
\includegraphics[scale=0.75]{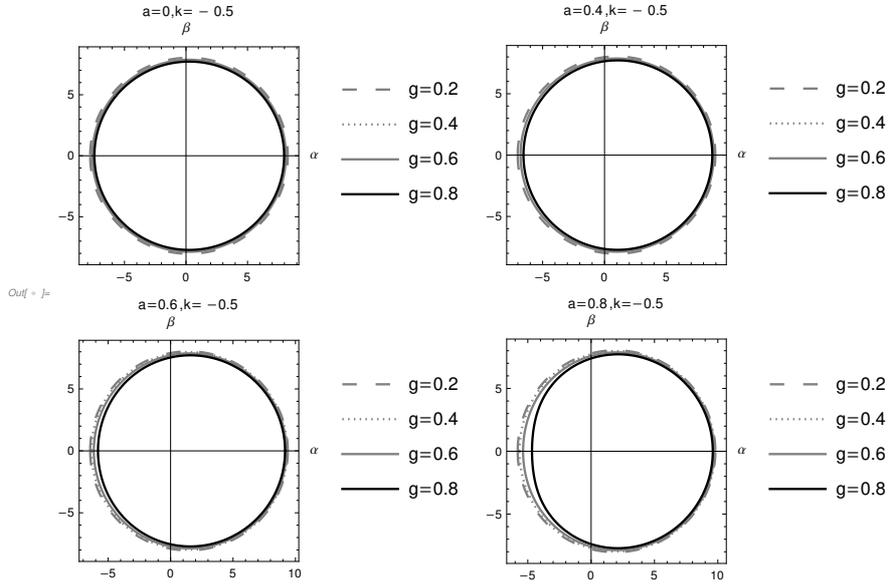}
\caption{shadow of gravitationally collapsed magnetic monopole in PFDM for varying magnetic moment g with different dark matter parameter k and spin parameter a}
\end{figure}

The shape of the blackhole can be charecterized by introducing two observables, considering three points $(\alpha_t,\beta_t)$,$(\alpha_r,\beta_r)$,$(\alpha_r,0)$ representing the tonp, bottom and extreme right respectively. the first observable $R_s$ is the radius of the shadow.Now let $D_s$ be the difference between left end and distortion point.the second observable $\delta_s$ is the distortion parameter and given as
\begin{equation}
\delta_s = \frac{D_s}{R_s}
\end{equation}
and radius of the shadow is given as
\begin{equation}
R_s = \frac{(\alpha_t -\alpha_r)^2-\beta_t}{2|\alpha_r-\alpha_t|}
\end{equation}
where the $\alpha_r$ and $\alpha_t$ can be obtained by setting $\beta = 0$ and $\partial_r \beta =0$. Figure 4 shows the evolution of shadow radius with respect to magnetic moment (g) for varying spin parameter with constant dark matter parameter, It is clear that the radius of the shadow decreases with increasing magnetic moment.

In The below fig [4] for a constant magnetic charge g, the radius of the shadow decreases with an increase in dark matter parameter k, On the contrary from Fig [5] the distortion parameter increases with an increase in dark matter parameter k. Although not prominent, the distortion parameter increase with an increase in the spin of the black hole. We got a similar result when varying magnetic charge with constant k, the radius decreases with increase in magnetic charge g, the distortion parameter increases with increase in g, the peculiarity here from the previous case is that there is a significant increase in distortion parameter as spin increases.

\begin{figure}[h]
\centering
\includegraphics{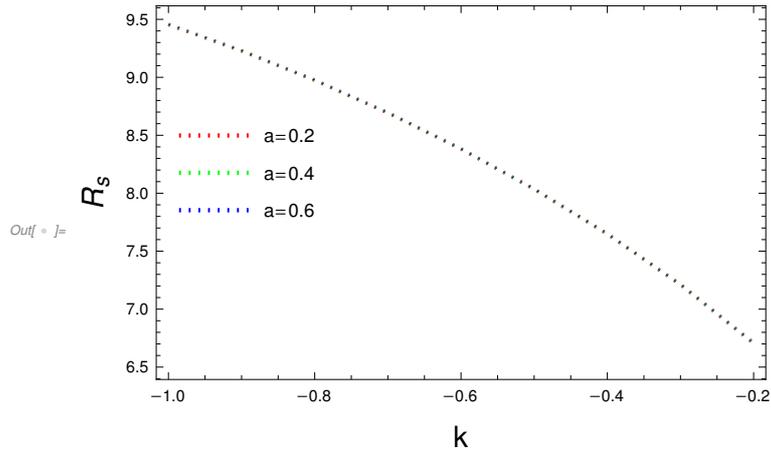}
\caption{Evolution of $R_s$ with k for g = 0.2 with varying spin parameter}

\end{figure}
\begin{figure}[h]

\centering
\includegraphics{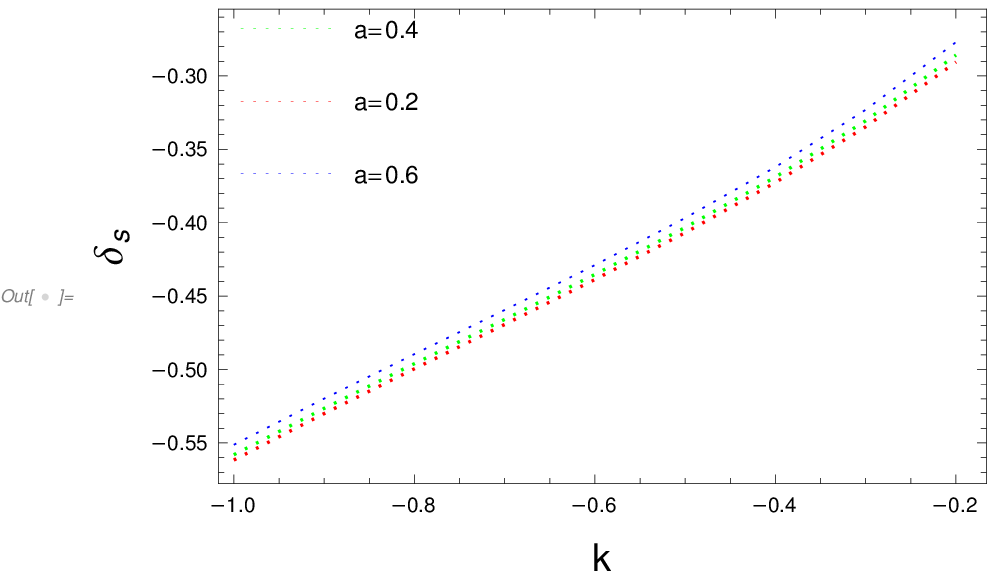}
\caption{Evolution of $\delta_s$ with k for g= 0.2 with varying spin parameter}
\end{figure}
\begin{figure}[h]
\centering
\includegraphics{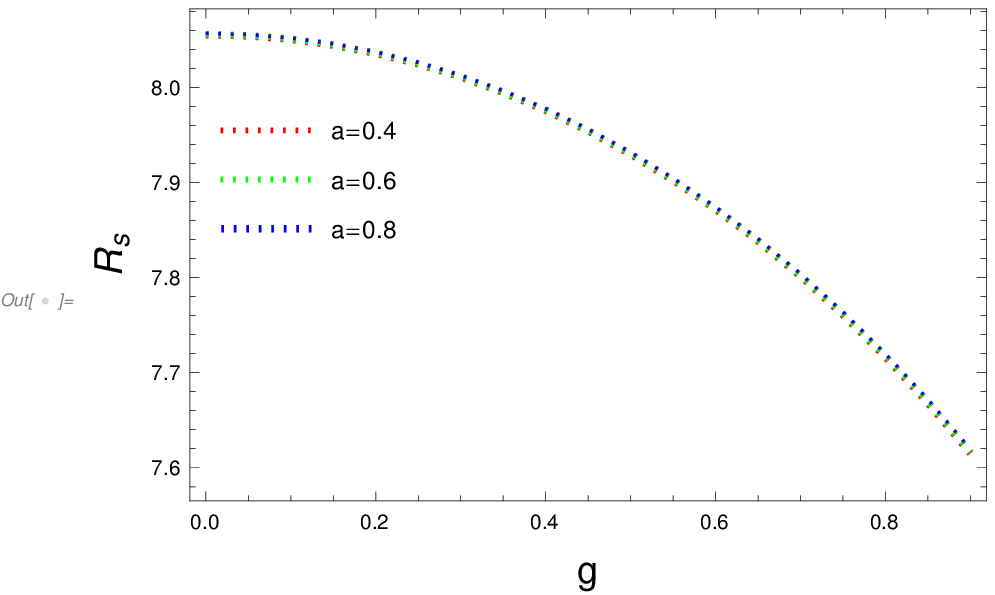}
\caption{Evolution of $R_s$ with g for k=-0.5 with varying spin parameter}

\centering
\includegraphics{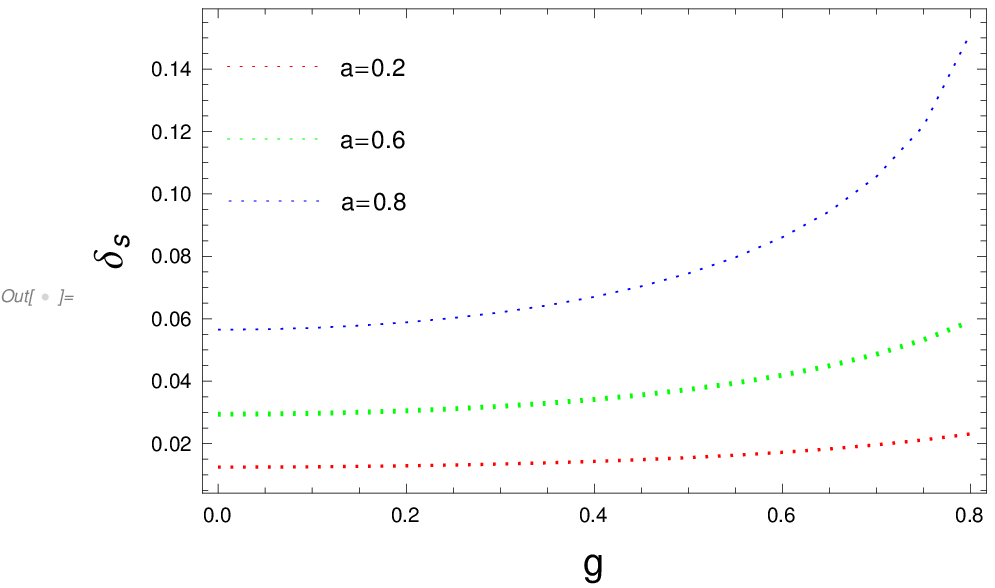}
\caption{Evolution of $\delta_s$ with g for k = -0.5 with varying spin parameter}
\end{figure}
\section{ENERGY EMISSION RATE}
In this sextion,we compute the energy emission rate of a rotating bardeen blackhole in perfect fluid dark matter. the expression for the energy emmision rate is given as,
\begin{equation}
\frac{d^2 E(\omega)}{d\omega dt} = \frac{2\pi^2 \sigma_{lim}}{e^{\omega/T}-1}\omega^3
\end{equation}
where $\omega$ is the frequency of photon and the limiting constant $\sigma_{lim}$ is
\begin{equation}
\sigma_{lim} \approx \pi R^2_s
\end{equation}
and Hawking temperature T is defined as,
\begin{equation}
T =\lim_{\theta=0,r\to r_+} \frac{\partial_r \sqrt{-g_{tt}}}{2 \pi \sqrt{g_{rr}}}
\end{equation}
where $r_+$ is the outer event horizon.

The energy emission rate concerning the frequency of photon has been plotted in Fig 6 and Fig 7. In Fig.6 For g=0.5, the emission rate increases with an increase in dark matter parameter and decreases with an increase in spin parameter. In Fig 7, for K = - 0.5, the emission rate decreases with an increase in magnetic charge and spin.
\newpage
\begin{figure}[h]
\centering
\includegraphics[scale=1.1]{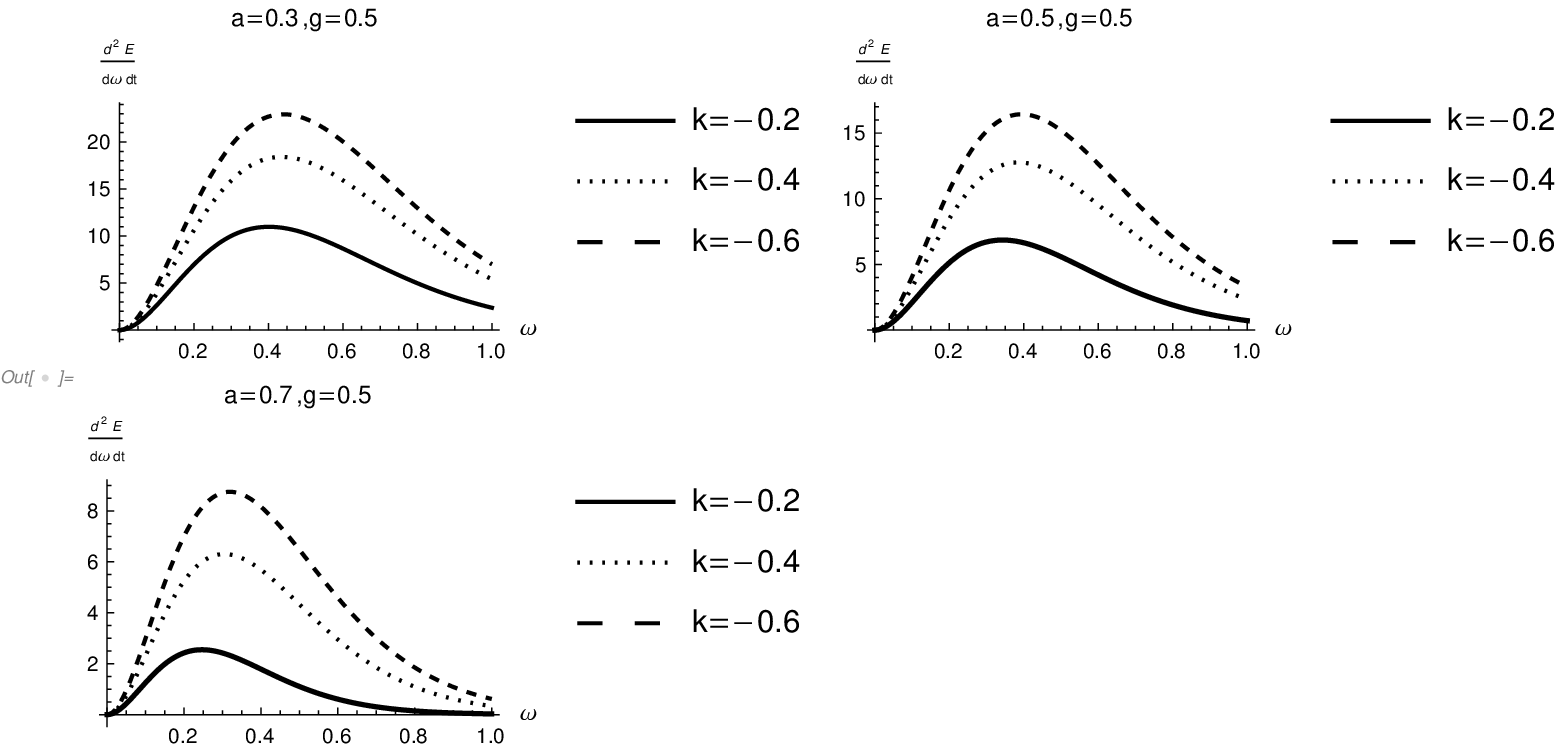}
\caption{Energy emission rate for g= 0.5 and different values of a and k}

\centering
\includegraphics[scale =1.1]{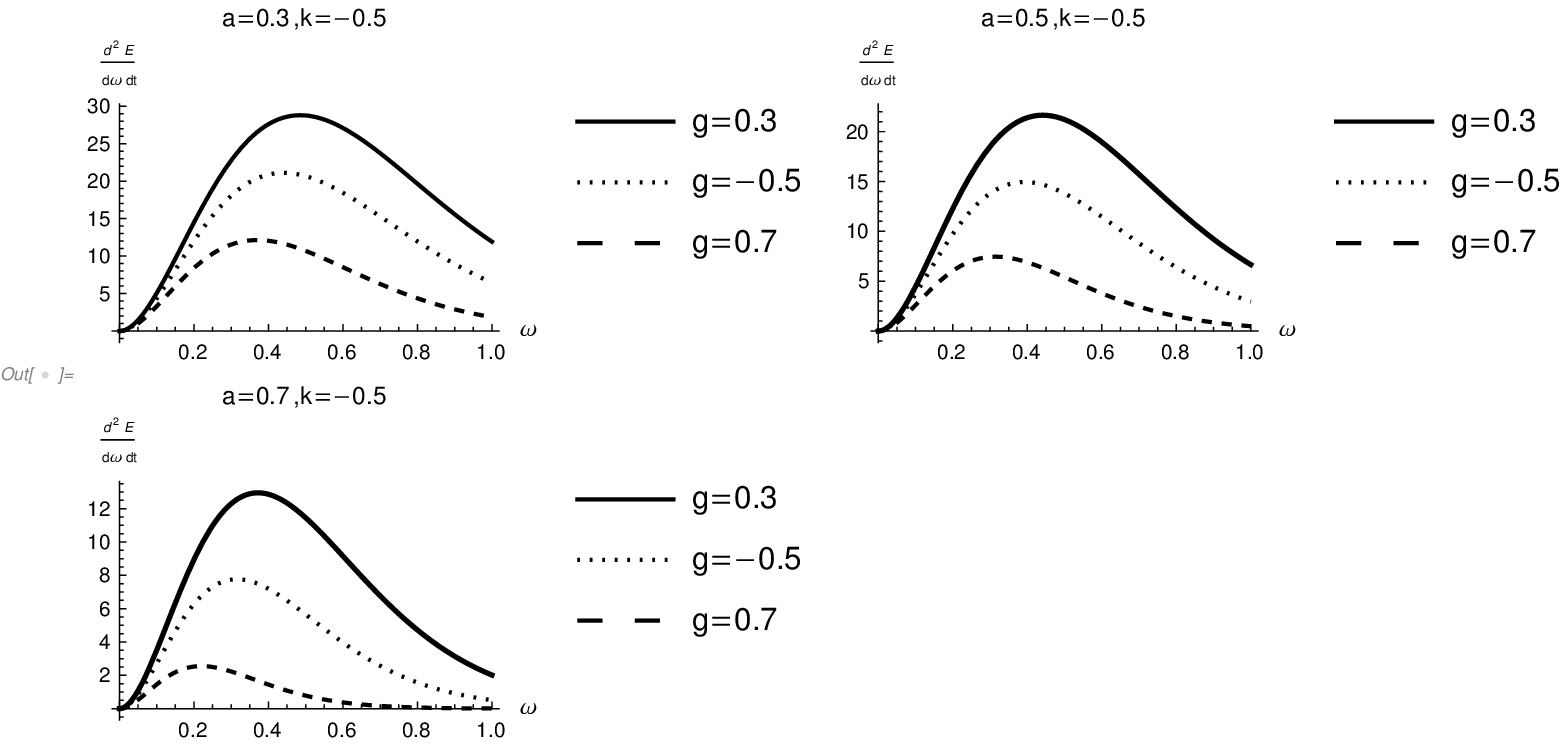}
\caption{Energy emission rate for k = -0.5 and different values of a and g}

\end{figure}

\section{Conclusion}

In this paper, we have studied the size and shape of the rotating Bardeen black hole shadow in presence of perfect fluid dark matter. we have discussed how parameters such a spin, magnetic monopole charge, and influence of dark matter affect the shadow of our black hole. we have found that for a constant magnetic moment g, the size of shadow decreases with an increase in dark matter parameter k and the distortion in the shape of shadow increases with an increase in spin parameter a
For a constant dark matter parameter k, Although there is no significant change in shadow size, the distortion caused by spin parameter a is prominent with an increase in g.the apparent shape of the black hole was studied by using two observables, the radius $R_s$ and the distortion parameter $\delta_s$. Further the blackhole emission rate is also studied, we found out that for rotating Bardeen in PFDM, For g=0.5, the emission rate increases with an increase in dark matter parameter and decreases with an increase in spin parameter. In Fig 7, for K = - 0.5, the emission rate decreases with an increase in magnetic charge and spin.

\bibliographystyle{unsrt}
\bibliography{draft}

\end{document}